\documentclass[a4paper]{article}
\usepackage[latin1]{inputenc}
\usepackage{graphicx}
\usepackage{amsmath}
\usepackage[numbers]{natbib}
\begin{document}
 
\baselineskip 18pt

\begin{center}
{\Large {\bf  Exotic atoms }}
 \vskip5mm Paul Indelicato$^{1,}$\footnote{Email address: paul.indelicato@spectro.jussieu.fr},
 \vskip3mm \mbox{}%
$^{1}$ Laboratoire Kastler-Brossel, \'Ecole Normale Sup\'erieure et
Universit\'e P. et M. Curie, Case 74, 4 place Jussieu, F-75252, Cedex 05,
France

\bigskip

\begin{abstract}
In this paper I review a number of recent results in the field of
exotic atoms.  Recent experiments or ongoing experiments with muonic,
pionic and antiprotonic hydrogen, as well as recent measurement of the
pion mass are described. These experiments provide information about
nucleon-pion or nucleon-antinucleon interaction as well as information
on the proton structure (charge or magnetic moment distribution).

\end{abstract}
\end{center}
\newpage

\section{Introduction}
\label{sec:intro}
Exotic atoms are atoms that have captured a long-lived, heavy
particle. This particle can be a lepton, sensitive only to the
electromagnetic and weak interactions, like the electron or the muon,
or a meson like the pion, or a baryon like the antiproton. An other
kind of exotic atom is the one in which the \emph{nucleus} has been
replaced by a positron (positronium, an e$^+$e$^-$ bound system) or a positively
charged muon (muonium, a $\mu^+$e$^-$ bound system). Positronium and muonium are pure quantum
electrodynamics (QED) systems as they are made of elementary,
point-like Dirac particles insensitive to the strong interaction. The
annihilation of positronium has been a benchmark of bound-state QED
(BSQED) for many years. For a long time there has been an outstanding
discrepancy between the calculated (see, e.g., \cite{afs2002} for a
recent review) and measured lifetime of ortho-positronium in vacuum,
that has been resolved recently \cite{vzg2003}. As the positronium is
the best QED test system, such a discrepancy was considered very
serious.  The $1 \,^3\mbox{S}_1 \to 2\, ^3\mbox{S}_1$ transition has also been measured
accurately \cite{fcmc93}.

Muonium has also been investigated in detail (see e.g., \cite{hug92}.)
Both the ground state hyperfine structure \cite{lbdd99} and the $1s
\to 2s$ transition \cite{cmyn88,dfcs89,sbbb95,mbbb2000} have been
investigated theoretically \cite{mky2001,hill2001,egs2001} and
experimentally. The work on the hyperfine structure provides a very
accurate muon mass value as well as a value for the fine structure
constant (see the CODATA recommended values of the fundamental
constants \cite{mat2000}).

The capture of a negatively charged, heavy particle $X^-$ by an atom,
occurs at a principal quantum number $n\approx n_e
\sqrt{\frac{m_{X^{-}}}{m_e}}$ where $n_e$ is the principal quantum number of
the atom outer shell, and $m_e$, $m_{X^-}$ are respectively the
electron and particle mass.  This leads to $n=14$, 16 and 41 for
muons, pions and antiprotons respectively. The capture process populates
$\ell$ sub-states more or less statistically. During the capture process
of an heavy, negatively charged particle, many or all of the electrons
of the initial atoms are ejected by Auger effect. As long as electrons
are present, Auger transition rates are very large and photon emission
is mostly suppressed except for the low lying states. For light elements, or particles like the
antiproton, the cascade can end up with an hydrogenlike ion, with only
the exotic particle bound to the nucleus \cite{sabb94}.

\begin{figure} [!hbp]
\includegraphics[clip=true,scale=0.5,angle=-90]{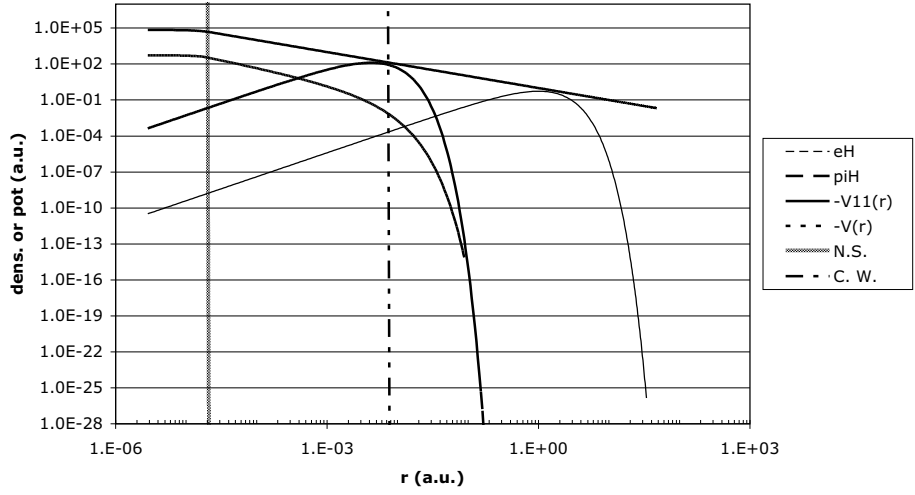}
\bigskip
\caption{Natural scales comparison in pionic and normal hydrogen.
 C.W: compton wavelength of the electron (QED scale). N.S.: proton radius (strong interaction scale).
 eH: hydrogen 1s electronic densisty. piH: pionic hydrogen 1s density.
 -V11: Uelhing (vacuum polarization) potential. V: Coulomb potential}\label{fig:elpidens}
\end{figure}

The spectroscopy of exotic atoms has been used as a tool for the study
of particles and fundamental properties for a long time. Exotic atoms
are also interesting objects as they enable to probe aspects of atomic
structure that are quantitatively different from what can be studied
in electronic or ``normal'' atoms. For example, all captured particles
are much heavier than the electron, and thus closer to the nucleus,
leading to a domination of vacuum polarization effects over
self-energy contributions, in contrast to normal atoms.  The different
relevant scales, Coulomb and vacuum polarization potential, together
with pionic and electronic densities in pionic and normal hydrogen are
represented on Fig.~\ref{fig:elpidens}. On can see that the pion
density inside the nucleus, or where the vacuum polarization
potential is large, is several orders of magnitude larger than the
electronic density in hydrogen. This lead to very large vacuum
polarization and finite nuclear size corrections.

Other fundamental changes can be found in exotic atoms: pions are
bosons, and thus obey the Klein-Gordon equation, while electrons,
muons and antiprotons as spin $1/2$ fermions, obey the Dirac
equation. Yet antiprotons, which are not elementary particles, have a
magnetic moment very different from the one of a Dirac particle. This
leads to large corrections not present in other types of atoms.

In the present paper, I will review a number of systems of interest
for the study of the proton structure or of the strong interaction at
low energy. In Section~\ref{sec:pi-muh}, I describe an ongoing experiment 
to measure the 2s Lamb-shift of muonic hydrogen. In section~\ref{sec:pi-piat}, I present
 recent experiments involving pionic atoms. Finally in section~\ref{sec:pi-pbar},
I will review recent work on antiprotonic hydrogen and helium.

\section{Muonic hydrogen and the determination of the proton charge radius}
\label{sec:pi-muh}
\begin{figure} [!hbp]
\includegraphics[clip=true,scale=1.0,angle=0]{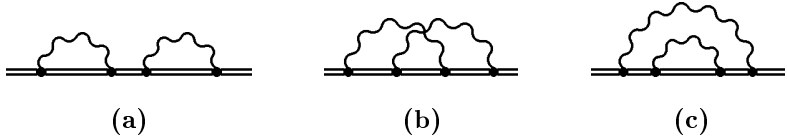}
\bigskip
\caption{The three Feynman diagrams contributing to the two-loop
self-energy. Double lines represent bound electron propagators and
wavy lines photon propagators. Diagram (a) represents the
loop-after-loop term. The irreducible part is obtained when the
propagator between the two-loop has an energy different from the
energy of the bound state being studied \protect\cite{yis2003}. \label{fig:sese-diag}}
\end{figure}

In the last decade very important progress has been made in the
accuracy of optical measurements in hydrogen. With the help of
frequency combs and rubidium fountain atomic clocks, the accuracy of the
1s$\to$2s transition measurement has reached $1.8\times 10^{-14}$,
giving 2~466~061~413~187~103(46)~Hz \cite{nhrp2000}. The Rydberg
constant (which requires knowledge of e.g., 2s$\to$nd transitions) that is needed to
extract the Lamb-shift from the 1s$\to$2s transition energy and to get
theoretical values is now known to $7.7\times 10^{-12}$ \cite{dsaj2000}. From this
information one can obtain the 1s and 2s Lamb-shift to 2.7~ppm accuracy.  However for many years it has been impossible to use
those very accurate measurements to test QED in hydrogen, which is the
only atom, at the present moment, in which experimental uncertainty is
much smaller than the size of two-loop BSQED corrections. Because in hydrogen
$Z\alpha << 1$, calculations of radiative corrections have been done as
an expansion in $Z\alpha$, i.e., expanding the electron propagator in
the number of interaction with the nucleus. It is only recently that for
the one-loop self-energy an exact, all order calculation, with a numerical 
precision small compared to the 46~Hz experimental error bar has been
performed. For the two-loop self-energy, the situation is very
complex. In the first calculation of the irreducible contribution to
the loop-after-loop contribution (Fig.~\ref{fig:sese-diag}~a) it was
found that the all-order contribution did not match, even at $Z=1$, the
result obtained by summing up all know term in the $Z\alpha$ expansion
\cite{mas98,mas98a}. This results was latter confirmed
\cite{yer2000}. It should be noted that this piece has no meaning by
itself, not being a renormalizable, gauge invariant set of Feynman
graphs.

\begin{figure} [!hbp]
\includegraphics[clip=true,scale=0.5,angle=0]{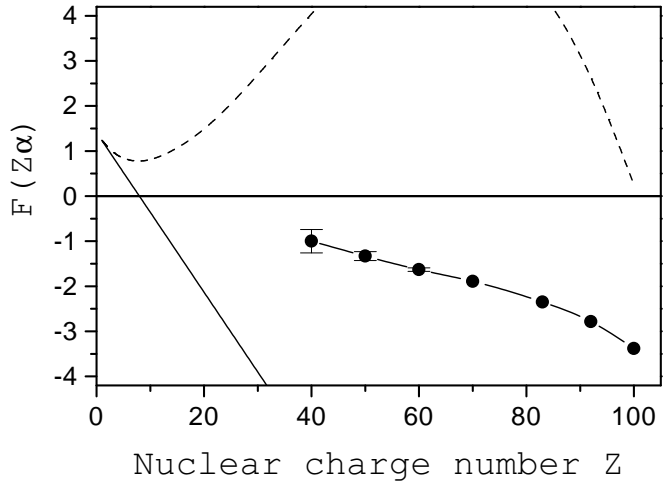}
\bigskip
\caption{Comparison with all-order numerical calculation and the function obtained from the first or second order
expansion in $Z\alpha$. \label{fig:sese-res}}
\end{figure}

More recently the complete all-order two-loop self-energy has been
evaluated, but only for $Z\ge 40$ \cite{yis2003}. It cannot be said at
the moment wether the extrapolation to $Z=1$ agrees with the $Z\alpha$
expansion (Fig.~\ref{fig:sese-res}).

Yet the issue cannot be resolved with the help of experiment, as the
proton charge radius is not well known, and the uncertainty on the
theoretical value due to that fact is larger than any uncertainty on
the two-loop corrections. Values of the proton radius measured by
electron scattering range from 0.805(12)~fm\cite{hand63} to 0.847(11)
\cite{mmd96} and 0.880(15)~fm \cite{ros2000}, the two later values
resulting from reanalysis of the same experiment\cite{simon80}. On the
other hand, one obtains 0.908~fm from a comparison between measurements
in hydrogen and QED calculation\cite{mat2000}.

Owing to this large dispersion of results, the uncertainty in QED calculations is 
4 times larger than the present experimental accuracy. It has thus been proposed to use
muonic hydrogen to obtain an independent measurement
of the proton radius. The experiment consists in measuring the 2s$\to$2p$_{3/2}$ transition energy,
which is strongly dependent in the proton radius. From
 value of \cite{egs2001} for the ``light by light'' contribution one gets
\begin{equation}
  \label{eq:muptote}
  206.099(20) - 5.2256 r^2 + 0.0363 r^3 \, \textstyle{\textrm{meV}},
\end{equation}
where the number in parenthesis represent the uncertainty (quadratic
sum), and where $r$, the proton mean spherical charge radius, must be
expressed in Fermi. If one uses instead Ref.~\cite{pac96,pac99},
then one obtains 206.074(3)~meV for the constant term.

An experiment aiming at an accuracy of 40~ppm of the 2s$\to$2p$_{3/2}$ energy difference
has been started at the Paul Scherrer Institute (PSI). Such an
accuracy would provide a proton charge radius to $\approx$~0.1\%
accuracy, which would allow to compare theory and experiment for the
1s and 2s Lamb-shift on hydrogen to the 0.4~ppm level. The experiment
uses the fact that the 2s state is metastable in muonic hydrogen. This
is due to the fact that, the muon being 206 times more massive than
the electron (and thus 200 times closer to the nucleus), vacuum
polarization dominates radiative corrections in exotic atoms, and has
opposite sign compare to self-energy. The 2s state is thus the lowest $n=2$
level in muonic hydrogen, while it is 2p$_{1/2}$ in hydrogen. The
experiment thus consists in exciting the 6~$\mu$m $2 \,^3\mbox{S}_{1/2} \to 2\, ^{5}\mbox{P}_{3/2}$
transition with a laser, and detects the 2~keV X-rays resulting from
the 2p$_{3/2}\to$1s transition that follows. In order to reduce
background events, coincidence between the high-energy electrons
resulting from the muon disintegration and the 2~keV X-rays must be done.

\begin{figure} [!hbp]
\includegraphics[clip=true,scale=0.5,angle=-90]{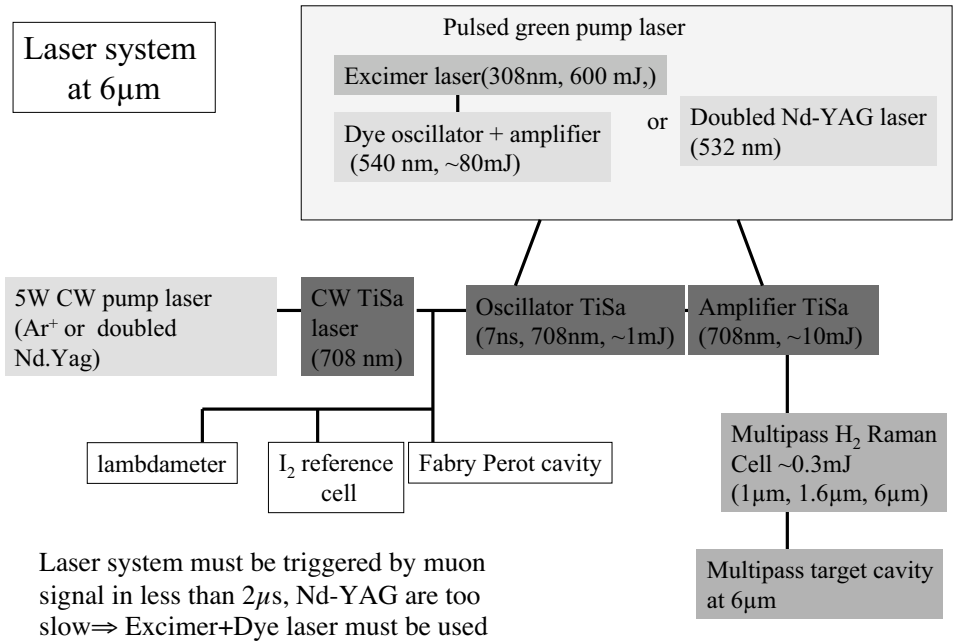}
\bigskip
\caption{Principle of the 6$\mu$m laser for the study of the muonic hydrogen 2s$\to$2p transition}\label{fig:laser6}
\end{figure}

The experiment uses slow muons prepared in the \emph{cyclotron trap
II} \cite{sim93}, installed on the high-intensity pion beam at PSI. The muons, originating
in pion decays, are decelerated to eV energies through interaction
with thin foils inside the trap. They are then accelerated to a few
keV, and transfered to a low density hydrogen target in a 5~T magnetic
field, using a bent magnetic channel, to get rid of unwanted
particles\cite{dhhk93,taq96}. A stack of foils at the entrance of the
target is used to trigger an excimer laser in around 1~$\mu$s (the
muons half life is around 2~$\mu$s). This laser is at the top of a
laser chain that use dye lasers, Ti-sapphire lasers and a multipass
Raman cell filled with 15~bars of H$_2$ to produce the 6~$\mu$m
radiation.  The laser system is shown on Fig.~\ref{fig:laser6}. More
details on the population of metastable states and on the experimental
set-up as can be found in \cite{tbch99,pbcd2000,kabc2001}.  A first
run of the experiment took place in summer 2002, in which an intensity
of 0.3~mJ was obtained at 6~$\mu$m, which is enough to saturate the
transition, and $\approx$~50 muons/h where detected in the
target. Counting rate is expected to be around 5 events/h at the
transition peak, which makes the experiment extremely difficult, owing
to the uncertainty on where to look for the transition, and of the complexity of
the apparatus.

\section{Pionic atoms}
\label{sec:pi-piat}
Pions are mesons, i.e., particles made of a quark-antiquark
pair. They are sensitive to strong interaction.  To a large extent,
the strong interaction between nucleons in atomic nuclei results from
pion exchange. The lifetime of the charged pion is $2.8\times
10^{-8}$~s. They decay into a muon and a muonic neutrino. The mass of
the pion is 273 times larger than the electron mass. Contrary to the
electron, it has a charge radius of 0.8~fm and it is a spin-0 boson.

\subsection{Measurement of the pion mass}
\label{sec:pi-pimass}
For a long time the spectroscopy of pionic atoms has been the favored
way of measuring accurately the pion mass. This mass was measured in 1986 in
pionic magnesium, with a crystal X-ray spectrometer, to a 3~ppm
accuracy \cite{jnbc86,jbde86}. Yet, as the pions were stopped in
solid magnesium, in which it was possible for the pionic atom to
recapture electrons before de-excitation, it was found that the
hypothesis made by Jeckelmann \emph{et al.} on the number of electron
recaptured in the atom was incorrect (in the pion cascade leading to
the formation of the pionic atom, all the electron are ejected by
auger effect.) This happened in experiments designed to measure the
muonic neutrino mass, from the decay of stopped positively charged pions into muon and
muonic neutrino \cite{abdf94}. This experiment found a negative value
for the square of the neutrino mass, using the 1986 value of the pion
mass.  A reanalysis of the 1986 experiment, with better modeling of the
electron capture was done, which lead to a pion mass value in agreement
compatible with a positive value for the square of the neutrino mass.

\begin{figure} [!hbp]
\includegraphics[clip=true,scale=0.5,angle=0]{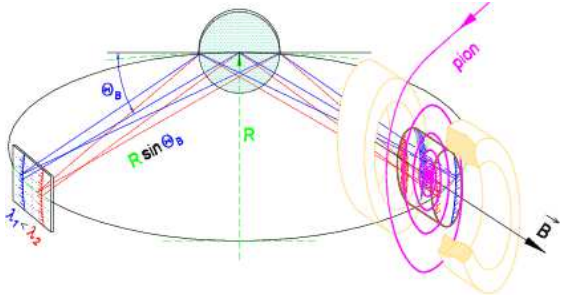}
\bigskip
\caption{Principle of X-ray spectrocopy of exotic atoms with the
cyclotron trap and a spherically curved crystal spectrometer. The
bidimensional X-ray detector is a 6-chips cooled CCD detector and is
located on the Rowland circle of radius $R/2$, where $R$ is the radius
of curvature of the crystal ($\approx 3$~m)
}\label{fig:trap}
\end{figure}

Such a situation was very unsatisfactory and it was decided to use the
cyclotron trap and the high-luminosity X-ray spectrometer, developed
initially for work with antiprotons at the Low Energy Antiproton Ring
(LEAR) at CERN to redo a measurement of the pion mass, in a low
pressure gas, in which case electron recapture is negligible
small. Moreover the resolution of the spectrometer was such that line
resulting from the decay of an exotic atom with an extra electron
would be separated from the main transition in a purely hydrogenlike
exotic atom. In a first experiment a value from the pion mass was
obtained by doing a measurement in pionic nitrogen, using copper
K$\alpha$ X-rays as a reference \cite{lbgg98}. This measurement, with
an accuracy of 4~ppm, and in good agreement with the limits set by
\cite{abdf94}, allowed to settle the question of the pion
mass. However the 4~ppm accuracy is not good enough for recent projects
involving pionic hydrogen, which are discussed in Sec. \ref{sec:pi-pih}.
The schematic of the experiment is shown in Fig.~\ref{fig:trap}.

The previous experiment accuracy was limited by the beam intensity,
the characteristics of the cyclotron trap, the quality of the X-ray
standard (broad line observed in second order of diffraction, while
the pion line was observed in first order) and the CCD size and
operation.  It was decided to use the 5g$\to$4f transition in \emph{muonic}
oxygen transition as a standard, with and energy close to the one of
the 5g$\to$4f transition in pionic nitrogen in place of the Cu energy
standard, relying on the fact that the muon mass is well known
\cite{mat2000}.  This standard energy can be evaluated with a
uncertainty of $\approx 0.3$~ppm.  A new trap was designed, optimized
for muon production, which was also to be used for the experiment
presented in Sec.~\ref{sec:pi-muh}.  Meanwhile the beam intensity of
the PSI accelerator had improved. Finally a new CCD detector was
designed, with larger size, higher efficiency and improved operations
\cite{naab2002}.

With these improvements, a new experiment was done, which lead to a
statistical accuracy in the comparison between the pionic and muonic
line, compatible with a final uncertainty around 1~ppm
\cite{naab2002a}. However at that accuracy, effects due to the
fabrication process of the CCD are no longer negligible and require
very delicate study, e.g., to measure the pixel size.  Those studies are
underway.

A byproduct of the pion mass measurement has been a very accurate
measurement of the 5$\to$4 transition fine structure in pionic
nitrogen. The energy difference between 5g$\to$4f and 5f$\to$4d is
found to be $2.3082\pm 0.0097$~eV \cite{lbgg98}, while a calculation
based on the Klein-Gordon equation, with all vacuum polarization
corrections of order $\alpha$ and $\alpha^2$ and recoil corrections
provides 2.3129~eV \cite{bpi2002}. This is one of the best test of QED
for spin-0 boson so far.

\subsection{3.2. Pionic hydrogen and deuterium}
\label{sec:pi-pih}
Quantum Chromodynamics is the theory of quarks and gluons, that
describe the strong interaction at a fundamental level, in the
Standard Model. It has been studied extensively at high-energy, in the
asymptotic freedom regime, in which perturbation theory in the
coupling constant can be used.  At low energy the QCD coupling
constant $\alpha_{\mbox{S}}$ is larger than one and perturbative
expansion in $\alpha_{\mbox{S}}$ cannot be done. Weinberg proposed
Chiral Perturbation Theory (ChPT) \cite{wei66} to deal with this
problem. More advanced calculations have been performed since then,
that require adequate testing. Short of the possibility of studying
pionium (a bound pion-antipion system) accurately enough, pionic
hydrogen is the best candidate for accurate test of ChPT. The shift
and width of np$\to$1s transition in pionic hydrogen due to strong
interaction can be connected respectively to the
$\pi^-$p$\to\pi^-$p and $\pi^-$p$\to\pi^0$n cross-sections,
which can be evaluated by ChPT, using a Deser-type
formula\cite{dgbt54}. After a successful attempt at studying pionic
deuterium with the apparatus described in Sec.~\ref{sec:pi-pimass},
which provided in a very short time a sizable improvement over
previous experiments \cite{hksb98}, it was decided that such an
apparatus could lead to improvements in pionic hydrogen of a factor 3
in the accuracy of the shift and of one order of magnitude in the
accuracy of the width. In order to reach such an improvement,
systematic studies as a function of target density and of the transition
(np$\to$1s, with n=2, 3 and 4) have been done.

\begin{figure} [!hbp]
\includegraphics[clip=true,scale=0.5,angle=-90]{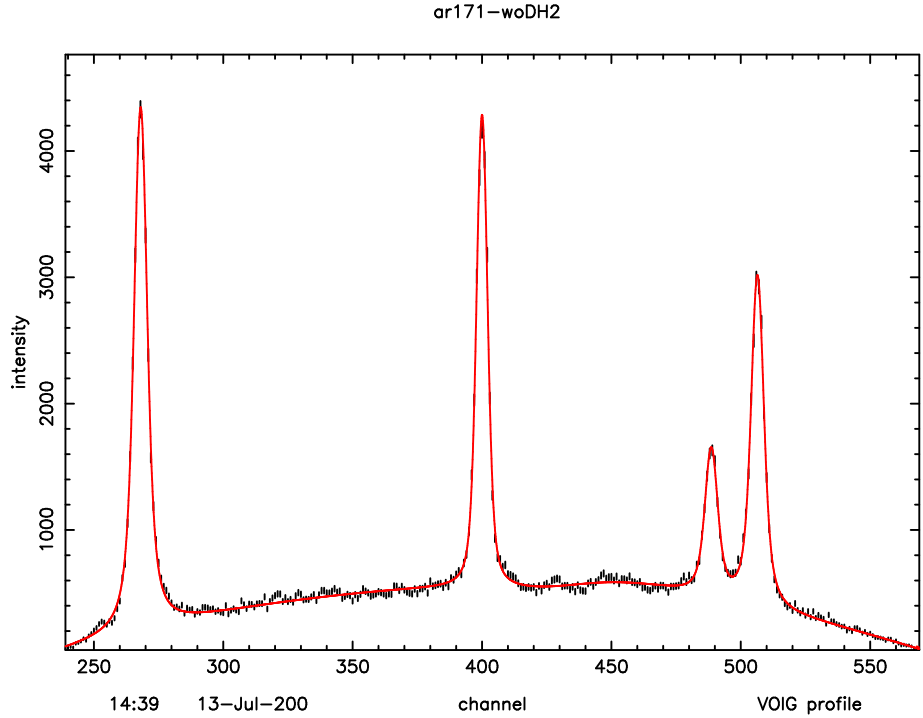}
\bigskip
\caption{High-statistic X-ray spectrum from the
1s2s$^3S_1\to$1s$^2\,^1S_0$ transition in helium-like argon (center
line) from the PSI Electron-Cyclotron Resonance Ion Trap. 37500 counts were accumulated in this line in
35~min, using the instrument in Fig.~\ref{fig:trap}.  The width of this
3.1~keV line is 0.4~eV. The line on the left is from Ar$^{14+}$ and
those on the right are from Ar$^{15+}$.
}\label{fig:ar16}
\end{figure}

The main difficulty in the experiment is to separate the strong
interaction broadening of the pionic hydrogen lines, from other
contributions, namely the instrumental response function, Doppler
broadening due to non-radiative de-excitation of pionic hydrogen
atoms by collisions with the H$_2$ molecules of the gas target and
from possible transitions in exotic hydrogen molecules.  The
instrumental response is being studied using a transition in helium-like
ions\cite{abbd2003}, emitted by the plasma of an Electron-Cyclotron
Ion Trap (ECRIT) build at PSI\cite{bsh2000}.  High-intensity spectra, allow
for systematic study of instrumental response. Exotic atoms do not
provide as good a response function calibration as most line coming
from molecules are broadened by Doppler effect due to the Coulomb
explosion during the atom formation process \cite{sabg2000}, and as
the rate being much lower, the statistic is often not sufficient. An
example of an highly-charged argon spectrum in the energy range of interest is
presented on Fig.~\ref{fig:ar16}.

\section{Antiprotonic atoms}
\label{sec:pi-pbar}
The operation of LEAR, a low-energy antiproton storage ring with
stochastic and electron cooling at CERN from 1983 to 1996 has caused a
real revolution in antiproton physics. Numerous particle physics
experiments were conducted there, but also atomic physics experiments.
A number of the latter used antiprotonic atoms produced with the
cyclotron trap (from $\bar{\mbox{p}}$H to $\bar{\mbox{p}}$Xe). Others
used Penning trap to measure the antiproton/proton mass ratio to test
CPT invariance \cite{gkhh99}.  An other experiment was concerned
with precision laser spectroscopy of metastable states of the
He$^+\bar{\mbox{p}}$ system \cite{hmtm94}, the existence of which had been
discovered earlier at KEK \cite{inss91}.  This experiment is now being
continued with improved accuracy at LEAR successor, the AD
(Antiproton Decelerator). Compared with recent high-accuracy
three-body calculations with relativistic and QED corrections, these
experiments provide very good upper bounds to the charge and mass
differences between proton and antiproton, again testing CPT
invariance\cite{hehi2001}. More recently the hyperfine structure of
the $^3$He$^+\bar{\mbox{p}}$ atoms as been investigated
\cite{weis2002} and found in good agreement with theory \cite{bk98}.
Expected accuracy improvements in the new AD experiment ASACUSA,
should lead to even more interesting results \cite{ymhw2002}.

\subsection{Antiprotonic hydrogen and deuterium}
\label{sec:pi-pbarh}
X-ray spectroscopy of antiprotonic hydrogen and deuterium was
performed at LEAR to study the strong interaction between nucleon and
antinucleon at low energy.  First the use of solid state detectors
like Si-Li detectors, provided some information. The study of line
intensities provided estimates of the antiproton annihilation in the
2p state. The 2p$\to$1s transitions were observed. While the
transition energy is around 8.7~keV, the 1s broadening due to
annihilation is $1054\pm 65$ eV and the strong interaction shift is
$-712.5\pm 25.3$~eV.  Measuring such a broad line is very difficult as
many narrow contamination lines will be superimposed on it. Moreover those rather precise
 values are spin-averaged quantities that neglect the unknown 1s level splitting. More
recently the use of CCD detectors has allowed to improve the
$\bar{\mbox{p}}$H measurement \cite{aabc99a} and to make the first
observation of $\bar{\mbox{p}}$D line, which is even broader, due to
three body effects \cite{aabc99}.

\begin{figure} [!hbp]
\includegraphics[clip=true,scale=0.5,angle=90]{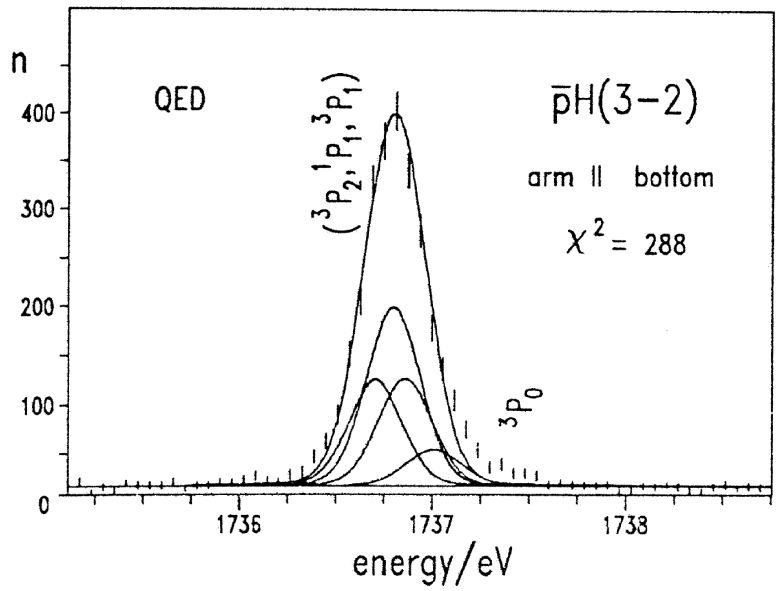}
\bigskip
\caption{High-resolution spectrum of antiprotonic hydrogen \cite{gaab99}. The difference between the measured line shape and the solid line is due to the fact
that the solid line represents a line-shape model without strong interaction, evaluated following the method described in the text \eqref{eqn:hfssec}. 
}\label{fig:pbar-spec}
\end{figure}

The broadening of the 2p state however is much smaller. The Balmer
3d$\to$2p lines can thus be studied by crystal spectroscopy. The use
of the cyclotron trap allowing to capture antiprotons in dilute gases
with a 90~\% efficiency and of an efficient, high-resolution
crystal spectrometer were instrumental to the success of such an
experiment, owing to the low production rate of antiprotons (a
few $10^8$ per hour). With that a counting rate of around 25 counts
per hour was observed, due to the use of an ultimate resolution device, even optimized for efficiency.
 The experimental set-up is basically the same
as described in Sec.~\ref{sec:pi-piat}. However to improve X-ray
collection efficiency, a double spectrometer was build, with two
arms symmetrical with respect to the trap axis and three crystals. On
one side a large CCD detector allowed to have two vertically
super-imposed crystals. On the other side a single crystal was
mounted. Resolutions of around 290~meV where achieved, which are of
the order of the expected line splitting. The final spectrum observed
with one of the three crystal/detector combination is presented on
Fig.~\ref{fig:pbar-spec}.

In order to extract strong interaction parameters from such spectrum,
a detailed description of the QED structure of the multiplet is
required.  Antiproton being composed of three quarks are not
point-like, Dirac particles.  In particular their gyro-magnetic ratio
is very different : for the antiproton
we have $a_{\bar{p}}=(g-2)/2 = -1.792847386$ instead of $-\alpha/(2\pi)\approx -1.2\times 10^{-3}$ for a
positron. The
corrections due to the anomalous magnetic moment of the antiproton can be accounted for
by introducing the operator (valid for distances
larger than the Compton wavelength of the electron $\hbar/mc)$
\begin{equation}
        \Delta H= a \frac{\hbar q}{2 m_p} \beta \left( i \frac{
        \boldsymbol{\alpha}\boldsymbol{E}}{c}- \boldsymbol{\Sigma} \boldsymbol{B} \right),
        \label{eqn:hfssec}
\end{equation}
where  $m_p$
is the antiproton mass, $\boldsymbol{E}$ and $\boldsymbol{B}$ are the
electric and magnetic fields generated by the nucleus,
$\boldsymbol{\alpha}$ are Dirac matrices and
\begin{equation}
\boldsymbol{\Sigma}=\left(\begin{array}{cc}
 \boldsymbol{\sigma}&0\\0& \boldsymbol{\sigma}
\end{array}
\right) .
\end{equation}

\begin{figure} [!hbp]
\includegraphics[clip=true,scale=0.5,angle=90]{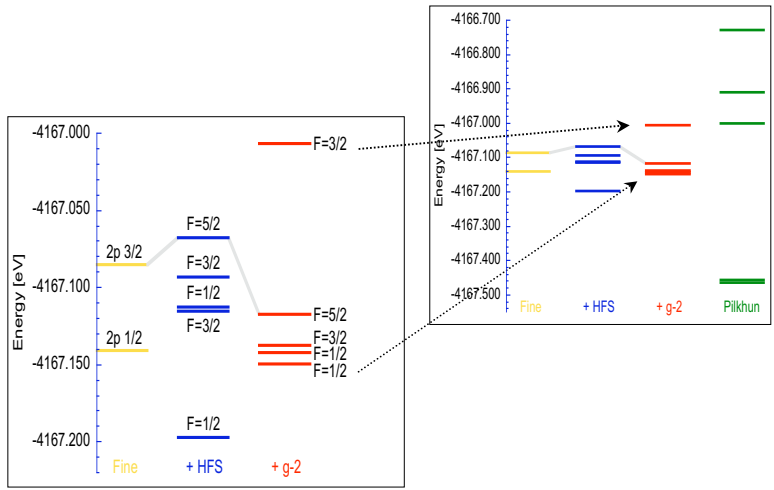}
\bigskip
\caption{Theoretical level scheme of antiprotonic deuterium and comparison with earlier work (Pilkuhn) \protect\cite{bps81}.
}\label{fig:pbar-deut}
\end{figure}

Vacuum polarization corrections and finite particle size (both for the
nucleus and the antiproton) must be included.  Because the antiproton
is $\approx 2000$ times closer to the nucleus than in the hydrogen
case, hyperfine structure and the $g-2$ correction are very large,
even larger than the fine structure. In such a case perturbation
theory is insufficient to account for the effect. The full Hamiltonian
matrix over the 2p and 3d Dirac states must be build and
diagonalized. The result of such a fully relativistic calculation for
antiprotonic deuterium is shown in Fig.~\ref{fig:pbar-deut}, together
with the result of earlier calculation \cite{bps81}. The large
difference between the two results is not understood.  However the
results from \cite{bps81} do not reproduce correctly the observed line
shape \cite{gaab99}.  By combining the high-precision measurements
carried out with the cyclotron trap and the spherical crystal
spectrometer, and the theoretical calculations presented above, it has
been possible to evaluate strong interaction shifts for the 2p level,
and different $\bar{\mbox{p}}$H and $\bar{\mbox{p}}$D spin states
\cite{gaab99}.  The results confirm calculations based on the
Dover-Richard semi-phenomenological potential \cite{dar80,ras82} (see
\cite{gaab99} for more details).


\section{Conclusion and perspectives}
In this paper, I have explored different aspects of the physics of
light muonic, pionic and antiprotonic atoms. I have left out many
aspects of that physics, like the study of the atomic cascade,
collisions between exotic atoms and gases, or the atomic and molecular
phenomena involved in muon-catalyzed fusion. The formation of
antiprotonic highly-charged ions as has been observed at LEAR seems to
point to very exciting new atomic physics\cite{rgfis2003}. I have not
explored either the studies of interest to nuclear physics like
measurement of nuclear charge distribution for heavier elements with
muonic atoms (see, e.g., \cite{bfgr89}) or of neutron distribution with
antiprotonic atoms\cite{tjlh201}. With the continuous progress in
accelerators technology (improvements in the intensity at PSI), the
development of very low energy antiproton energy at the AD at CERN,
the trapping of antihydrogen\cite{aabb2002,gbos2002}, or the new
antiproton machine at GSI, it is expected that this physics will
continue to develop in the years to come and provide more challenges
to atomic and fundamental physics.

\bigskip\noindent
{\large \bf Acknowledgments}
 Laboratoire Kastler Brossel is Unit{\'e} Mixte de Recherche du CNRS
 n$^{\circ}$ C8552.  I wish to thank all the participants to the
 antiprotonic hydrogen, pion mass, pionic hydrogen and muonic hydrogen
 experiments for their relentless effort to make those experiments live
 and develop. I am in particular indebted to D. Gotta, L. Simons,
 F. Kottmann and F. Nez. On the theoretical side, participation of
S. Boucard, V. Yerokhin, E.O. Le~Bigot, V. Shabaev and P.J. Mohr is gratefully
acknowledged.

\bibliography{exotic}

\newpage


\end{document}